\documentclass[prd, 11pt, a4paper, showpacs, notitlepage, dvipsnames, reprint, twocolumn, preprintnumbers, nofootinbib, floatfix]{revtex4-1}

\usepackage{geometry}
\usepackage[utf8x]{inputenc}
\usepackage{makecell}
\usepackage{float}
\usepackage{tabularx}
\usepackage[activate={true,nocompatibility},final,tracking=true,kerning=true,factor=1500,stretch=10,shrink=10]{microtype}
\usepackage{graphicx}
\usepackage{bm, amssymb, amsmath, amsfonts}
\usepackage{multirow}
\usepackage{xcolor}

\usepackage{url}

\usepackage{url}
\usepackage[breaklinks]{hyperref}

\usepackage{hyperref}
\hypersetup{
colorlinks = true, 	    
linkcolor = magenta,	
citecolor = blue,		
}

\usepackage[linesnumbered,ruled,lined,boxed]{algorithm2e}
\usepackage[noend]{algpseudocode}

\SetAlCapSty{xAlCapSty}

\SetCommentSty{mycommfont}

\SetNlSty{mynlfont}{}{} 

\RestyleAlgo{algoruled}

\setcitestyle{numbers,square}

\begin{document}

\title{A machine learning algorithm for minute-long Burst searches} 
\author{Vincent Boudart$^{1}$}	\email[]{vboudart@uliege.be}
\author{Maxime Fays$^{1}$} \email[]{Maxime.Fays@uliege.be}

\affiliation{${}^1$ STAR Institute, Bâtiment B5, Université de Liège, Sart Tilman B4000 Liège, Belgium}

\begin{abstract}
\noindent   
Minute-long Gravitational Wave (GW) transients are events lasting from few to hundreds of seconds. In opposition to compact binary mergers, their GW signals cover a wide range of poorly understood astrophysical phenomena such as accretion disk instabilities and magnetar flares. The lack of accurate and rapidly generated gravitational-wave emission models prevents the use of matched filtering methods. Such events are thus probed through the template-free excess-power method, consisting in searching for a local excess of power in the time-frequency space correlated between detectors. The problem can be viewed as a search for high-value clustered pixels within an image, which has been generally tackled by deep learning algorithms such as Convolutional Neural Networks (CNNs). In this work, we use a CNN as a anomaly detection tool for the long-duration searches. We show that it can reach a pixel-wise detection despite trained with minimal assumptions, while being able to retrieve both astrophysical signals and noise transients originating from instrumental coupling within the detectors. We also note that our neural network can extrapolate and connect partially disjoint signal tracks in the time-frequency plane.
\end{abstract}
\maketitle 



\section*{Introduction}

The first gravitational wave (GW) event, coming from the coalescence of two black holes, was detected by the Advanced LIGO \cite{aLIGO} interferometers on September 14, 2015 \cite{first_detection}. Two years later, on August 17, 2017, the first binary neutron star (BNS) merger was observed \cite{neutron_star}, paving the way for the search for longer-duration gravitational waves. These two events are part of a broader class of signals called Compact Binary Coalescence (CBC), which further includes recently observed black hole - neutron star collisions \cite{NSBH}. To this date, only CBC events, coming from powerful sources and being accurately modelled systems, have been detected \cite{GWTC3}. However, new sources are expected to be observed given the planned sensitivity improvement of the Advanced LIGO and Advanced Virgo detectors \cite{Virgo}. Among the proposed candidates, unmodeled GW transients, known as bursts, cover a wide range of poorly understood astrophysical phenomena for which accurate waveforms are not available. Bursts include accretion-disk instabilities \cite{adi}, non-axisymmetric deformations in magnetars \cite{magnetars}, supernovae \cite{supernovae}, gamma-ray bursts \cite{GRB}, fallback accretion events \cite{fallback_accretion_NS}. The above-mentioned family of possible GW progenitor events contains both short- ($<2$ seconds) and long-duration signals (between 2 and several hundreds of seconds). In this paper, we present a novel machine learning algorithm targeted at Anomaly detection for Long-duration BUrst Searches (ALBUS). \\

Minute-long burst searches usually consist in finding an excess of power in the cross-correlated data of two or more detectors, known as correlated spectrograms or time-frequency (TF) maps. Algorithms that are searching for gravitational waves in these spectrograms can be classified in two categories: seed-based or seedless. Seed-based methods aim at clustering pixels above a predefined threshold while seedless algorithms are processing pixels derived from generic models. The current generation pipelines are the long-duration configuration of coherent WaveBurst (cWB) \cite{cWB}, the two different versions of the Stochastic Transient Analysis Multi-detector Pipeline - All Sky (STAMP-AS), Zebragard and Lonetrack \cite{STAMP_1,STAMP_2}, PySTAMPAS \cite{PySTAMPAS}, cocoA \cite{cocoA}, and X-SphRad \cite{Fays}. cWB, PySTAMPAS and X-SphRad are seed-based algorithms while the two STAMP-AS pipelines, Zebragard and Lonetrack, use seed-based and seedless clustering algorithms respectively. As well as Lonetrack, cocoA is a seedless pipeline. In opposition to CBC searches, no machine-learning based algorithm has yet been applied to long-duration searches. The uncertainties in the existing physical models of long-duration transients lead us to make minimal assumptions on progenitor sky-position, inclination, time-of-arrival and GW waveform characteristics. The waveform models cannot consequently be taken as accurate patterns to be recognised and are thus used as tests for algorithms and pipelines rather than actual targets of the search. The present work aims at circumventing this problem by mimicking template database of generic long-duration signals, allowing us to take advantage of the speed and the robustness of convolutional neural networks. \\

In section \ref{section:data}, we describe the data-generation method as well as the strategy to mimic long-duration signals. In section \ref{section:deep_learning}, details about the architecture and the training method are given. We present the results of the training and the detection performances both on injected signals and on background spectrograms in section \ref{section:results}. Section \ref{section:conclusion} is finally dedicated to discussions and conclusions.

\section{Data}\label{section:data}

\subsection{Time-Frequency map generation}

One of the detection methods of long-duration GWs is based on the excess of power method \cite{basics_GW}. It consists in correlating the output of a least two different detectors, under the assumption that their noise is uncorrelated. This produces coherent spectrograms where GW signals are represented as minute-long high-correlation patterns. The problem of detecting long-duration GWs can therefore be reduced to finding a cluster of pixels with high intensity in background noise. \\

The exact formulation of the coherence between two signals $x$ and $y$ is:
\begin{equation}
    C_{x\,y}(f) = \frac{|G_{x\,y}(f)|^2}{G_{x\,x}(f)\,G_{y\,y}(f)}
    \label{coherence}
\end{equation}

\noindent
where $G_{xy}(f)$ is the cross-spectral density (CSD) between signals $x$ and $y$, and $G_{xx}(f)$ and $G_{yy}(f)$ the power spectral density (PSD) of $x$ and $y$ respectively. The cross-spectral density is defined as:
\begin{equation}
    \begin{split}
    G_{x\,y}(f) = \int_{-\infty}^{\infty} \left[\lim_{T\to\infty} \frac 1 {T} \int_{-\infty}^{\infty} x_{T}(t-\tau)\,y_{T}(t) dt \right] &\\ e^{-i 2 \pi f \tau} d\tau
    \end{split}
    \label{cross spectral density}
\end{equation}

\begin{figure*}[!htb]
    \centering
    \includegraphics[trim={2.3cm 0.5cm 1cm 2cm}, clip, scale=0.56]{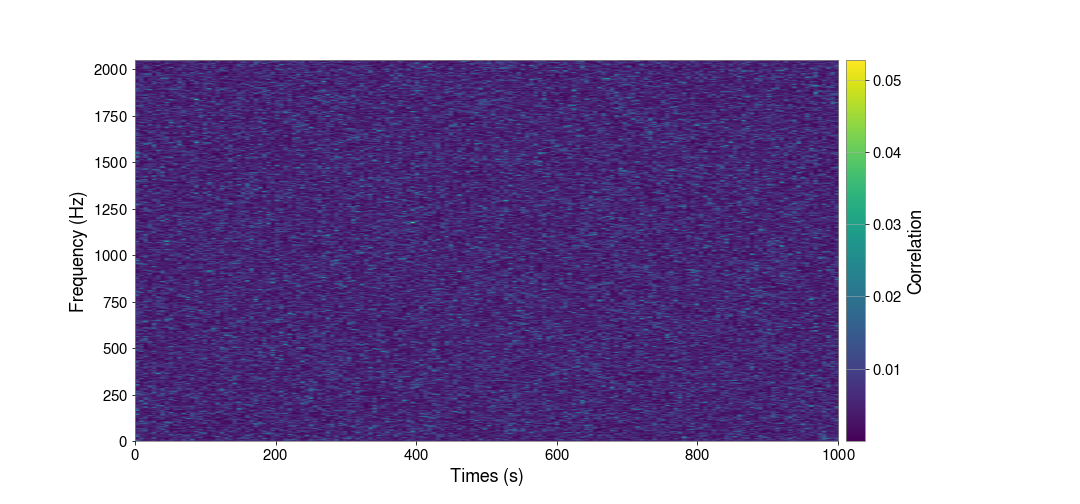}
    \caption{Typical time-frequency map of cross-correlated O3a background noise. The GPS time at the start of the Advanced LIGO Hanford (H1) and Advanced LIGO Livingston (L1) data is respectively 1246174396 and 1246108524.}
    \label{example TF map}
\end{figure*}

\noindent
$G_{xx}(f)$ and $G_{yy}(f)$ are then a particular case of (\ref{cross spectral density}) where respectively $y$ is replaced by $x$ and conversely. As the latter expression is intractable for a real signal, we employ Welch's method \cite{Welch} as an approximation method. Evaluating (\ref{coherence}) at several sampled frequencies then leads to the generation of a vector of coherence values versus frequency bins. To generate a full time-frequency array, we apply Welch's method to small subsets of our original signal. This is equivalent to updating the coherence value after some time and compiling this time evolution as a single map. The resulting spectrogram is finally whitened along the time axis by summing all the values of a frequency bin along the time axis and dividing by the total sum. An example of the obtained spectrogram is shown in Figure \ref{example TF map}. \\ 

To constitute a sufficient number of background spectrograms, we use time-slides \cite{time_slides} where the detector data are shifted by time delays larger than the time of flight of GWs between detectors. The time delays are also larger than the duration of target signals ($\geq$ 500s) to guarantee our cross-correlated data to contain only detector noise. \\ 









The time and frequency resolutions of the generated spectrograms have an impact on the sensitivity of the search. The longer the time segments, the more GW energy will be accumulated in a single pixel, leading to a higher coherence. As the noise appears to be coherent on very small time scales ($\ll$ 1 second) \cite{detchar2021}, increasing the length of the time segments allows to reduce the coherence of the noise versus the coherence of hypothetical signals, yielding an increased signal-to-noise ratio (SNR). However, longer time bins will cause short signals ($\sim10$ seconds) to fall into very few pixels, making them harder to detect. An identical reasoning can be made for the frequency resolution. In this work, we use a 6-second time resolution combined with a 2 Hertz frequency bin as a good compromise. Taking a 1000s data stream from the Advanced LIGO interferometers spanning frequencies up to 2048 Hz (see Figure \ref{example TF map}), results in spectrograms with dimensions of \textit{166}x\textit{1025} pixels. \\

\begin{figure*}[!htb]
    \centering
    \includegraphics[trim={1.5cm 0.5cm 0cm 1cm}, clip, scale=0.6]{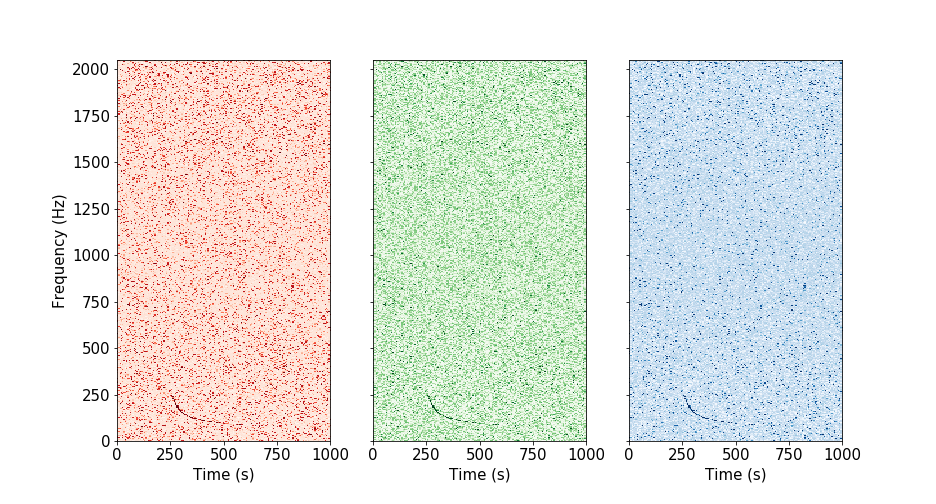}
    \caption{RGB channels of an 03a time-frequency background map where a \textit{GRBplateau} waveform \cite{GRBplateauShort} is injected. Note how the signal is visible in all 3 channels, from 250s to 550s. The GPS time at the start of the H1 and L1 data is respectively 1246174396 and 1246108524.}
    \label{example RGB}
\end{figure*}

The time-frequency arrays are heavy to store in the default 32-bit \textit{Python} precision because of their large amount ($>10^5$) of pixels. Each TF map weights about 13 megabytes. This substantial storage space can reduce the maximum batch size, i.e the number of training samples that are passed through the network in a single forward pass, accessible for the training, eventually increase the training time and downgrade the test performance \cite{bengio_recommendations}. It is also recommended to use batch sizes above 10 to avoid a highly noisy gradient descent, and to take advantage of the speed-up of matrix-matrix products over matrix-vector products \cite{bengio_recommendations}. A noisy gradient descent is obtained when the gradients obtained after each training iteration poorly generalizes to all the samples in the dataset and in fine extend the time needed for the network to converge. Saving the spectrograms as RGB images (8-bit integers) allows to reduce the storage needed at the cost of a small loss of precision in the values of the array. The maximum loss of precision for values falling right in between two integer levels is $(1/2^8)/2$, which is less than $0.2\%$. We thus have 3 channels displaying different information depending on the colormap used to draw the initial array. We choose the "cubehelix" colormap from the \textit{Matplotlib} \textit{Python} library \cite{matplotlib}, displaying the GW signals clearly in all 3 channels, as seen in Figure \ref{example RGB}. The final reduction factor in memory is roughly 26 compared to the 32-bit \textit{NumPy} arrays \cite{Numpy}, ultimately giving access to larger training batches. A further argument in favour of using RGB images is the wide use of the format in deep learning applications, e.g. some neural networks even require a 3-channel image as input \cite{YOLO, EfficientNet, CCSN}. \\

Therefore, the final dataset that we consider is made up of RGB time-frequency images generated from the coherence between the data from the Advanced LIGO Hanford (H1) and Advanced LIGO Livingston (L1) detectors, gathered during the first phase of the third advanced LIGO observing run (O3a).

\subsection{A new measure of excess power}

In order to form our dataset, we need to inject signals so that our neural network can learn to recognize their patterns in spectrograms. The current way of adding long-duration burst signals to background noise makes use of the root sum squared value of the strain $h(t)$ to gauge the strength of the injection. It is defined as :
\begin{equation}
    h_{rss} = \sqrt{\int_{}^{}\,h(t)^2\,dt}
    \label{hrss}
\end{equation}

\begin{figure}[!htb]
    \centering
    \includegraphics[trim={0.7cm 0.5cm 0cm 1cm}, clip, scale=0.42]{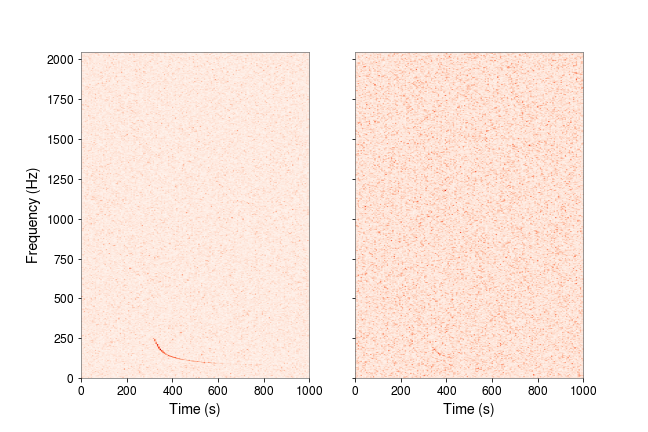}
    \caption{Injection of a \textit{GRBplateau} waveform from  \cite{GRBplateauShort} at 300s with a $h_{rss}$ value of $5\,10^{-22}$ in two different O3a background spectrograms. The GPS time at the start of the H1 and L1 data in the left panel is respectively 1246174396 and 1246108524, while for the right panel it is 1248305006 and 1248273184.}
    \label{hrss_problem}
\end{figure}

However, this expression only depends on the strain value of the injected signal, and makes the human eye visibility of the injection vary with the local noise level of the background map. Figure \ref{hrss_problem} shows a waveform model injected with a $h_{rss}$ value of $5\,10^{-22}$ in two different O3a background spectrograms. When the noise level is sufficiently high, as in the right panel of Figure \ref{hrss_problem}, the signal can be buried in the background noise, preventing a clear detection for the network. Such samples can fool a neural network during a supervised learning and eventually cause the training to be badly conditioned \cite{class_noise}. A new criterion is therefore required to form a healthy dataset. \\

To be visible in time-frequency representations, an injected signal has to stand out of the local noise level. Defining a noise-only spectrogram $N_{ij}$ and the same spectrogram to which a signal has been injected by $S_{ij}$, the new criterion is : 
\begin{equation}
    V = \sum_{i,j}\, \big( S_{ij}\,-\,N_{ij} \big)
    \label{visibility}
\end{equation}

\noindent
where the sum is carried over all the pixels $(i,j)$ in the map. We call this new criterion ``visibility". The pixel-to-pixel difference allows to fine tune how much signal emerges from the local noise level and form a dataset with different levels of intensity. This is particularly useful when training procedures like curriculum learning \cite{curriculum_learning} are used.

\subsection{Mimicking long-duration signals}

\begin{figure*}[!htb]
    \centering
    \includegraphics[trim={0cm 0cm 0cm 0cm}, clip, scale=0.35]{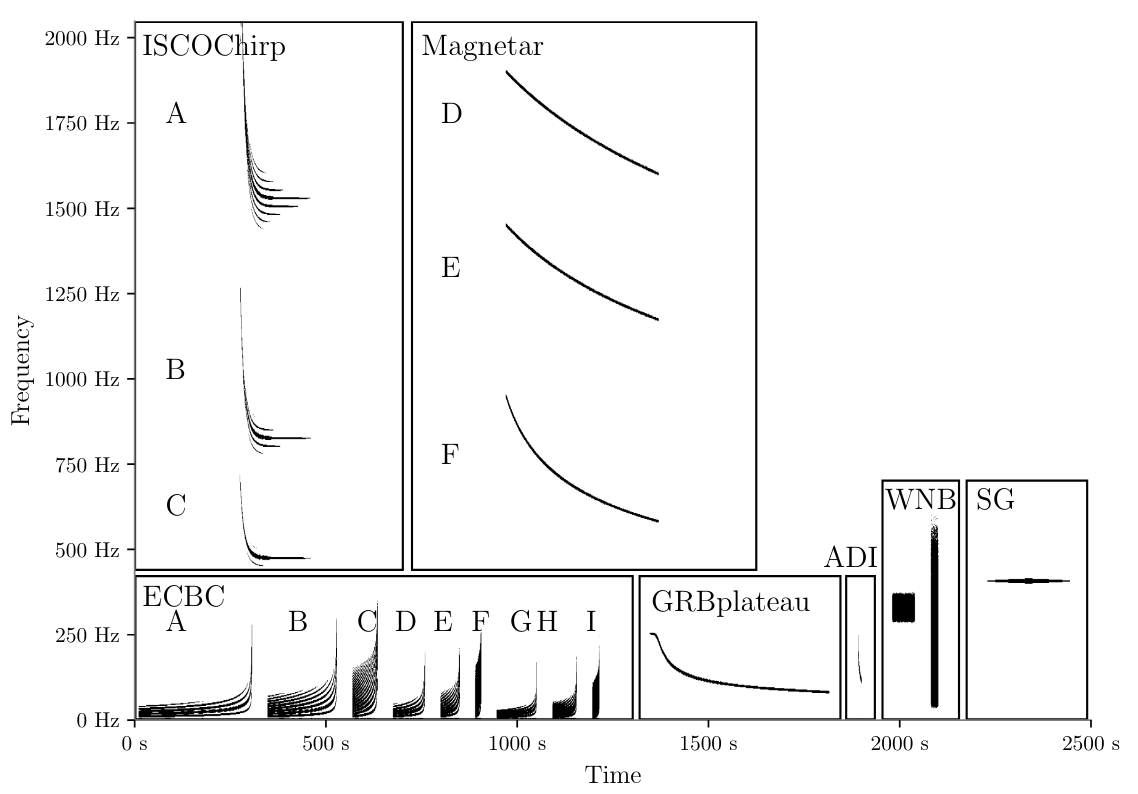}
    \caption{Signal models (waveforms) targeted by long-duration pipelines for the third observing run \cite{O3_long_duration_paper}.}
    \label{example waveforms}
\end{figure*}

\begin{figure}[!htb]
    \centering
    \includegraphics[trim={0.3cm 0cm 0cm 0cm}, clip, scale=0.52]{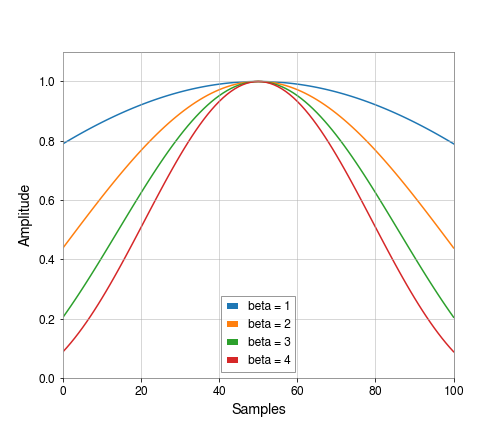}
    \caption{Examples of Kaiser windows with different values for the $\beta$ parameter.}
    \label{kaiser window}
\end{figure}

Neural networks are particularly good at recognising and classifying shapes and objects they have seen in training (YOLO \cite{YOLO}, AlexNet \cite{AlexNet}, GoogLeNet \cite{GoogleNet}, etc). They are therefore well-suited for detecting signals in time-frequency images if the training set is sufficiently close to the expected long-duration GWs. Figure \ref{example waveforms} shows the models used to test the long-duration pipelines for the third advanced LIGO-Virgo observing run. The waveforms generally show a \textit{chirp up} or \textit{chirp down} behavior. This property can be easily mimicked thanks to the \textit{Scipy Python} package \cite{scipy}. Specifically, it allows to draw time series with varying parameters such as the duration, the frequency bandwidth, and the frequency evolution. The frequency evolution can either be linear, hyperbolic, quadratic or logarithmic. Once the chirp signal is generated, the energy distribution is adjusted thanks to a Kaiser filter \cite{kaiser_window} as seen in Figure \ref{kaiser window}. For this, a window twice as long as the signal is generated and the chirp signal is multiplied by either the first or second half of the window. When the first half is implied, the final chirp shows a increasing energy distribution and the reverse holds when the second half is concerned. A shape parameter $\beta$ is used to control the width of the Kaiser window. A larger beta parameter implies a narrower Kaiser window leading to a more unbalanced energy distribution, the signal showing more intensity in the beginning or in the end of the chirp structure. \\

\begin{figure*}[htb]
    \centering
    \includegraphics[trim={0cm 0cm 0cm 0cm}, clip, scale=0.55]{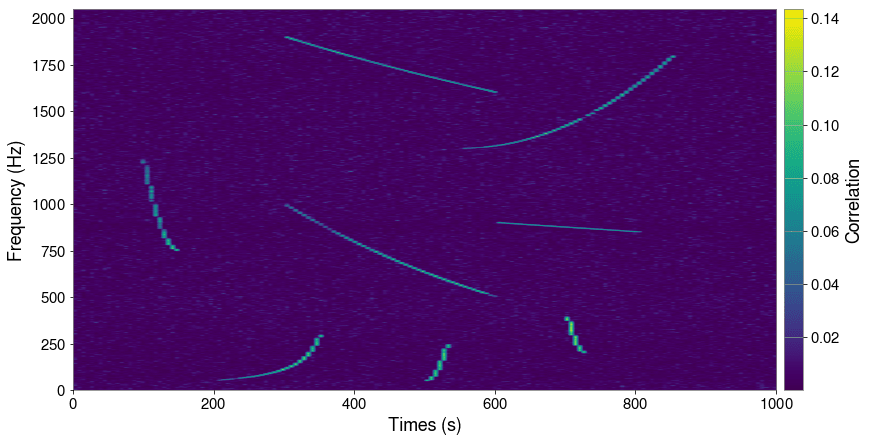}
    \caption{Examples of chirp signals produced for the training of our neural network.}
    \label{example chirps}
\end{figure*}

Figure \ref{example chirps} shows examples of chirp signals. These chirps are nearly identical to the models that have been selected to test long-duration algorithms. However, the harmonics that appear in GW emission models like \cite{ISCOchirp} or \cite{ecbc} cannot be reproduced with the Scipy library \cite{scipy}. These harmonics come from emission mechanisms such as multiple mass moments in the torus around black holes \cite{vanputten} or eccentricity oscillations in eccentric compact binary coalescences \cite{ecbc}. They usually show less power than the main component of the gravitational wave and show up exclusively for high amplitude injections, usually easily recognized in time-frequency images. We will thus make use of \cite{scipy} to generate our dataset and train our neural network without a-priori knowledge of the expected burst signals. \\

\section{Deep Learning approach}\label{section:deep_learning}
\subsection{Neural network in GW applications}

Neural networks and machine learning techniques have been recently applied to gravitational wave physics \cite{MLreview}. Among the variety of networks used, convolutional neural networks (CNNs) \cite{CNN}, being very good at pattern and shape detection \cite{YOLO,AlexNet}, have shown successful applications in the detection of black hole collisions \cite{CBC_cnn}, identification of the GW counterpart from supernovae \cite{Melissa}, binary neutron star detection \cite{Baltus, Krastev2020} and estimation of their parameters \cite{parameter_estimation}, as well as in classification of detector glitches \cite{GravitySpy}. CNNs have also been used in short burst detection with \cite{short_duration_burst}. \\

The strong capability of CNNs to recognize patterns can either be applied to one-dimensional (time series) \cite{CBC_cnn, Baltus, Krastev2020} or two-dimensional data \cite{GoogleNet, GravitySpy}. Their efficiency to detect shapes has even been adapted to Generative Adversarial Networks (GAN) \cite{Original_GAN}. Such networks have recently been used to generate short duration bursts \cite{CGAN_bursts}. \\

Given the performance, speed and wide application of machine learning techniques in the gravitational wave domain, we decide to apply it to the long duration search. Specifically, we use CNNs to detect and precisely localize chirp signals in the time-frequency space.

\subsection{ALBUS}

In order to detect signatures of long-duration bursts, we will also make use of CNNs. Most of the CNNs used to detect patterns and objects also involve a classification task \cite{YOLO, EfficientNet, AlexNet}. However, we want to highlight the pixels that could resemble to burst signals rather than assigning a label to the whole input map. We found that the convolutional network built in \cite{albus} returns a pixel-by-pixel localisation map. They associated a boolean target map to every image in the training set. The training method then consists in minimizing a loss between the output map and the target map, so that the former keeps approaching the latter as the training progresses. We will follow the same strategy but also bring some modifications, notably on the network architecture and on the definition of the target map. \\

\begin{figure*}[!htb]
    \centering
    \includegraphics[trim={0cm 0cm 0cm 0cm}, clip, scale=1.45]{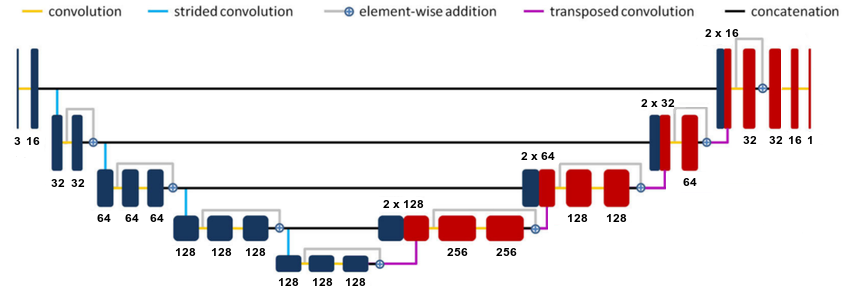}
    \caption{Architecture of ALBUS, modified from \cite{albus}. The downscaling and upscaling parts are represented in blue and red respectively. These two parts are coupled thanks to skipped connections, represented as concatenation lines. The numbers in black indicate the number of filters used at each stage of the network.}
    \label{architecture albus}
\end{figure*}

The network, shown in Figure \ref{architecture albus}, is made up of two parts, a downscaling part that keeps the useful information through its different layers, and an upscaling part that aims at localising precisely this information in a map with the same dimensions as the input of the network. The connections between the downscaling and upscaling parts help the network to learn how to precisely position the signals. The number of filters at every step, indicated below each layer in Figure \ref{architecture albus}, has been divided by 2 compared to \cite{albus}, reducing both the training time and the memory usage. \\

\begin{figure}[!htb]
    \centering
    \includegraphics[trim={0.7cm 0.5cm 0cm 1cm}, clip, scale=0.42]{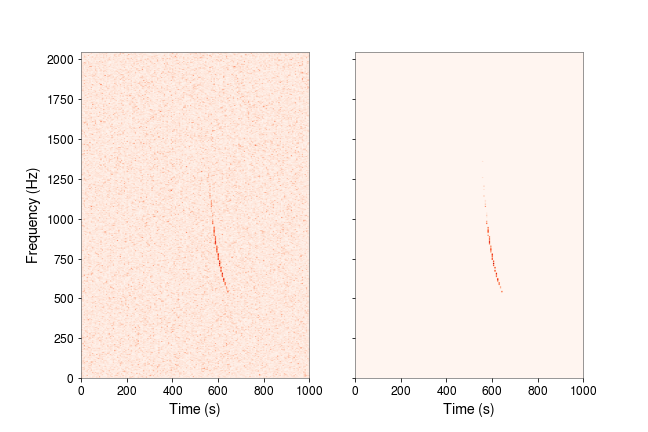}
    \caption{Background spectrogram in which a chirp signal is injected at 100s (left) and its associated target map used for the training phase (right). The energy distribution of the chirp signal can also be seen in the target map, providing adequate samples for the training phase.}
    \label{target map}
\end{figure}

The target map definition has also been modified. The localisation map alone is not sufficient to rank the spectrograms based on their content. As an example, a score can be defined as the sum of the pixels in the localisation map, which helps distinguishing GW candidates from noise-only images. If we keep the definition of \cite{albus}, all the pixels that will be highlighted by the network will be put to 1 in the localisation map. Therefore, summing up all the 1's in the image can lead to a high score, even if the pixels are scattered in the map. In such a case, it becomes harder to identify GW events through a unique score. 
We rather need a definition that can output both high and low values depending on the intensity of the signal injected in the image. We set a threshold on the spectrogram pixels corresponding to the $99^{th}$ percentile of the values. This is equivalent to keeping the top $1\%$ pixels showing the highest values. We then normalize the spectrogram. This procedure leads to a target map that follows the intensity evolution of the signals through the input map. An example of a spectrogram containing a chirp signal and its corresponding target map can be seen in Figure \ref{target map}. \\

The last adaptation concerns the loss used to train our neural network. The loss defined by \cite{albus} is a weighted mean squared error (MSE): 
\begin{equation}
    L = \frac{1}{2}\,\sum_{i,j}\, \big( T_{ij} + \lambda\,\overline{T}\big) \,\big( T_{ij}\,-\,O_{ij} \big)^{2}
    \label{loss Xing}
\end{equation}

\noindent
where $T$ and $O$ respectively stand for the target and output maps, $\overline{T}$ is the mean of the target map and $\lambda$ is a control parameter. However, our dataset is also composed of background images where no signal has been injected, leading to empty target maps. It is essential to add them in the training loop so that the network will not automatically try to find something in the input TF maps. With empty maps, the left parenthesis in \ref{loss Xing} is null, preventing the loss to give any feedback to the network in the backpropagation loop, which does not happen with the classical MSE loss, defined as :
\begin{equation}
    MSE = \frac{1}{2}\,\sum_{i,j}\,\big( T_{ij}\,-\,O_{ij} \big)^{2}
    \label{loss MSE}
\end{equation}

\noindent
As the MSE loss still gives a non-zero response with background images for which $T_{ij}$ is zero everywhere, we decided to choose it for the training of our neural network.

\section{Results}\label{section:results}
\subsection{Training ALBUS}

The dataset for the training is composed of 4500 background images and 4500 chirp images. The chirp signals were injected with 9 levels of \textit{visibility} (500 samples for each intensity). All the parameters for injecting chirp signals are summarized in Table \ref{parameters injection}. The validation set is made of $10\%$ of both the background and injection dataset. The delay indicates the time from the start of the spectrogram where signals are injected. We set a low frequency threshold at 30 Hz because of the high noise level of the Advanced LIGO detectors at lower frequencies \cite{detchar2021}. The chirp signals being drawn at a chosen $h_{rss}$ value, the visibility level is evaluated after every injection. An iterative loop then allows to obtain the desired visibility levels by adapting accordingly the initial $h_{rss}$ value. Because of this iterative loop, we tolerate a $\pm10\%$ range around the selected values to cover a wider space and to converge faster. \\

\begin{figure}[!htb]
    \centering
    \includegraphics[trim={0.5cm 0.3cm 1.5cm 1.0cm}, clip, scale=0.40]{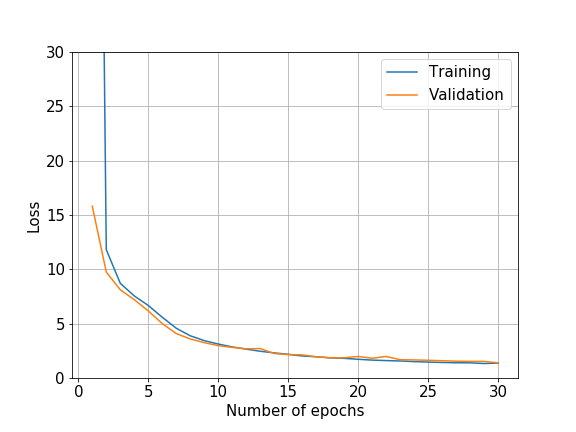}
    \caption{Training and validation losses for a 30-epoch training of ALBUS.}
    \label{loss training}
\end{figure}

The training algorithms have all been coded with \textit{PyTorch} \cite{Pytorch}. The ADAM optimizer \cite{Adam} has been chosen with a learning rate of $10^{-4}$. The batch size is set to 20 where one half is taken from the background images and the other half from the injection images. The training and validation losses for a training phase of 30 epochs are shown in Figure \ref{loss training}. The training loss decreases monotonically which suggests that the learning progresses evenly. The validation loss remains in close vicinity of the training loss ruling out any overfitting on the training data. We decided to stop the training after 30 epochs because both losses started to reach a plateau, indicating that no major improvements are made by the network. 

\begin{table}[!htb]
   \centering
   \renewcommand{\arraystretch}{1.5}
   \begin{tabular}{|c|c|}
     \hline
        & \textbf{Range of values} \\
     \hline
       \textbf{Duration} & 10-500 s \\
     \hline
       \textbf{Delay} & 0-500 s \\
     \hline
       \textbf{Frequency range} & 30-2000 Hz \\
     \hline
       \textbf{Frequency evolution} & lin. quad., log. or hyperbolic \\
     \hline
       \textbf{$\beta$ parameter} & 1-4 \\
     \hline
       \textbf{Visibility levels} & 12, 14, 16, 18, 20, 30, 40, 50, 60\\
     \hline
   \end{tabular}
   \caption{Parameters used to inject chirp signals in the TF maps. All the parameters are uniformly drawn from their range of values.}
   \label{parameters injection}
\end{table}

\subsection{Detection performance}

\begin{figure*}[!htb]
    \centering
    \begin{tabular}{cc}
        \includegraphics[scale=0.42, trim={0.5cm 0.5cm 1.5cm 1cm},clip]{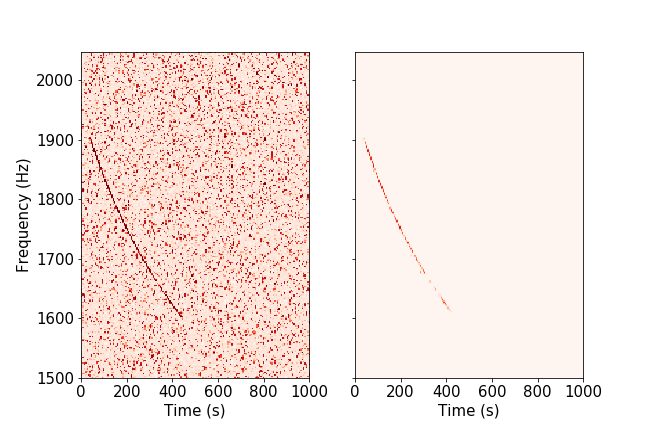} &
        \includegraphics[scale=0.42, trim={0.5cm 0.5cm 1.5cm 1cm},clip]{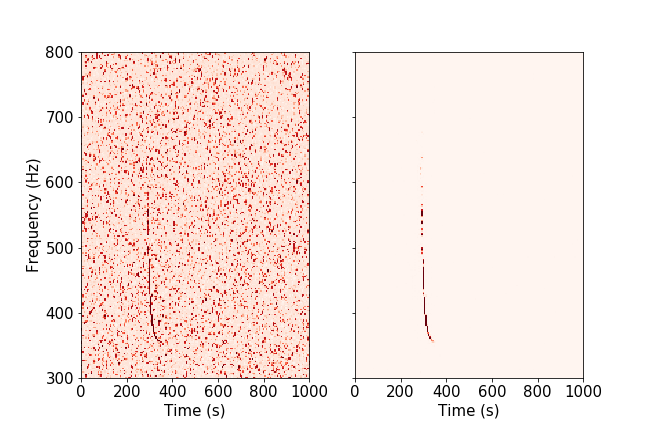}
    \end{tabular}
    \begin{tabular}{cc}
        \includegraphics[scale=0.42, trim={0.5cm 0.5cm 1.5cm 1cm},clip]{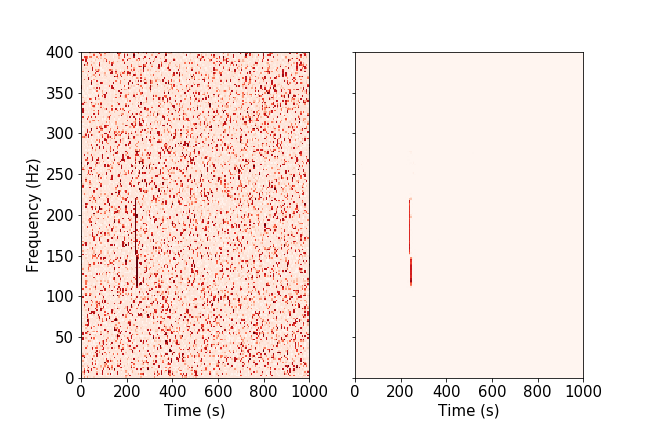} &
        \includegraphics[scale=0.42, trim={0.5cm 0.5cm 1.5cm 1cm},clip]{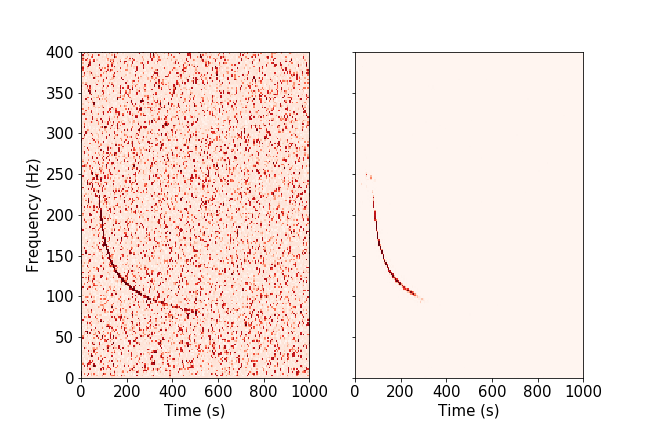}
    \end{tabular}
    \caption{Examples of detection performance on long-duration waveforms (\textbf{top left} : \textit{Magnetar-D} \cite{magnetar_waveform}, \textbf{top right} : \textit{ISCOchirp-C} \cite{ISCOchirp}, \textbf{bottom left} : \textit{ADI} \cite{adi} and \textbf{bottom right} : \textit{GRBplateau} \cite{GRBplateauShort}). The left image of each panel is the red channel of the input image and the right panel shows the output of ALBUS.}
    \label{examples detection}
\end{figure*}

Figure \ref{examples detection} shows the output of ALBUS for 4 different waveforms from the selected models in \cite{O3_long_duration_paper}. The simulated signals are well recognized and the variation of intensity in the input map is also seen in the localisation map. \\

An additional remark can be made concerning the upper right panel of Figure \ref{examples detection}, where a few pixels above the curve (around 600 Hz and 300 seconds) are highlighted in the output map. This behaviour is also observed in the lower right panel. Indeed, our network is not only looking at the pixels having a high value but also at the connectivity between these pixels. It then naturally looks prolongs the main structure to catch pixels following the general trend of the signal. Such a propriety can be a relevant tool to reject background images showing isolated hot pixels. \\

\begin{figure}[!htb]
    \centering
    \includegraphics[trim={0.5cm 0.5cm 1.5cm 0.5cm}, clip, scale=0.42]{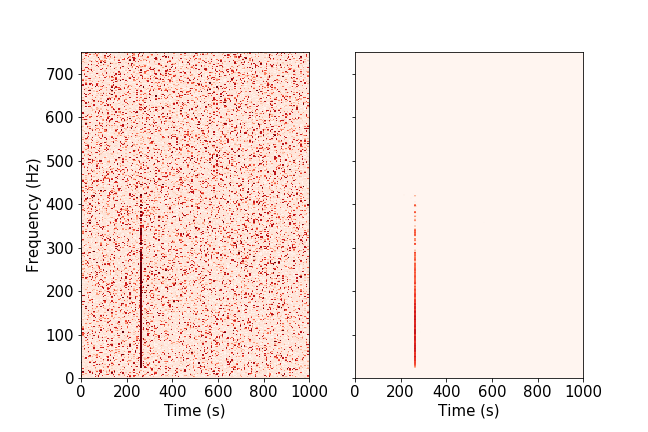}
    \caption{Example of glitch detection. The left image shows the red channel of the input map while the right panel shows the output of ALBUS.}
    \label{example glitch}
\end{figure}

Another detection capability is observed with transient noises called glitches. Glitches are appearing in the detector data in abundant quantities, produced due to several sources such as instruments or the environment \cite{noise_characterization, noise_characterization2}. Several classes of these artefacts have been identified through machine learning algorithms \cite{GravitySpy} and all show particular time-frequency morphologies. As most of the glitches last less than 6 seconds \cite{detchar2021}, they show up in our TF maps as straight vertical lines. The cross-correlation method reduces their impact since a glitch from L1 data needs to fall in the same pixel as another glitch from H1 data to show up in the correlated TF map. That small number of cross-correlated glitches in the background can explain why ALBUS does not consider them as part of the background noise and actually detects them. Figure \ref{example glitch} shows an example of a glitch and its localisation by ALBUS. \\

\begin{figure}[!htb]
    \centering
    \includegraphics[trim={0.5cm 0.5cm 1.5cm 0.5cm}, clip, scale=0.42]{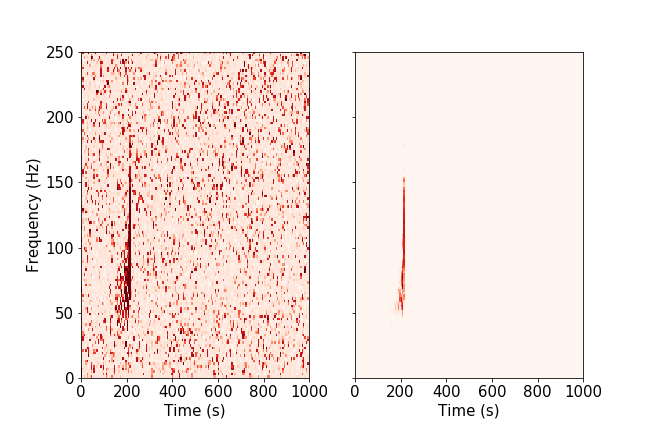}
    \caption{Example of waveform showing harmonics (\textit{ECBC-C}). The left image shows the red channel of the input map while the right panel shows the output of ALBUS.}
    \label{example harmonics}
\end{figure}

The final point concerns harmonics of waveform models. It is reasonable to foresee that ALBUS will not detect them as they are not showing up in the training dataset. Figure \ref{example harmonics} displays this phenomenon to some extent. The harmonics that appear to the left of the rising chirp are not found in the output map except in the very beginning of the signal around 50 Hz. This effect is certainly due to the extrapolation capability mentioned above, for which ALBUS is "looking" for smooth connected signals. The choice of not incorporating harmonics of chirps in the data has minimal consequence since the waveform models showing these harmonics are still detected. \\

\begin{figure*}[!htb]
    \centering
    \begin{tabular}{c}
        \includegraphics[scale=0.53, trim={2.3cm 0.5cm 1.5cm 1cm},clip]{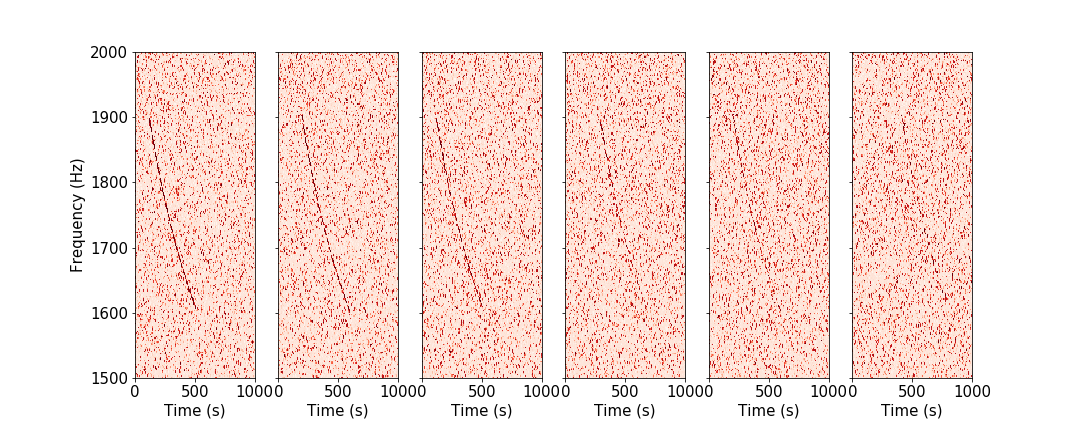}
    \end{tabular}
    \begin{tabular}{c}
        \includegraphics[scale=0.53, trim={2.3cm 0.5cm 1.5cm 1cm},clip]{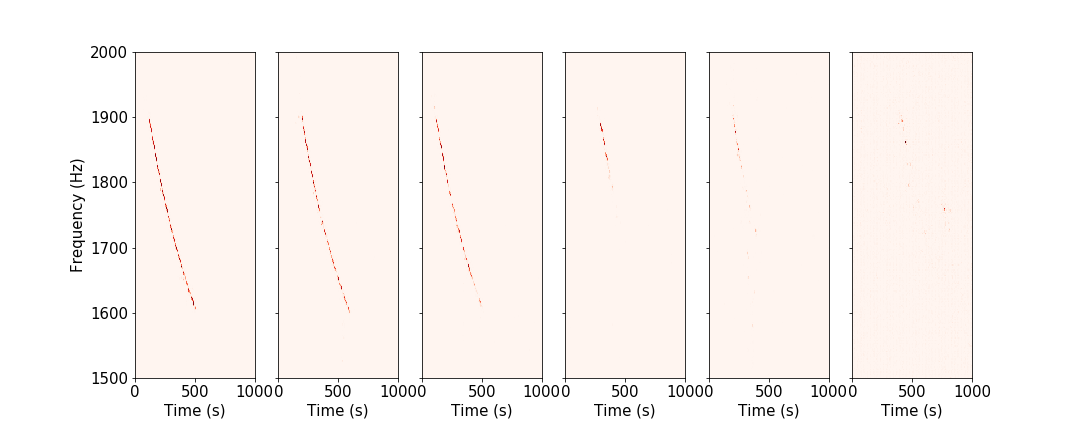}
    \end{tabular}
    \caption{Detection performance of ALBUS on 6 different visibility levels (from left to right : 60, 40, 30, 20, 16 and 12) for the waveform model \textit{Magnetar-D}. The top panel shows the input images and the lower panel shows the output of ALBUS.}
    \label{detection visibility 1}
\end{figure*}

\begin{figure*}[!htb]
    \centering
    \begin{tabular}{c}
        \includegraphics[scale=0.53, trim={2.3cm 0.5cm 1.5cm 1cm},clip]{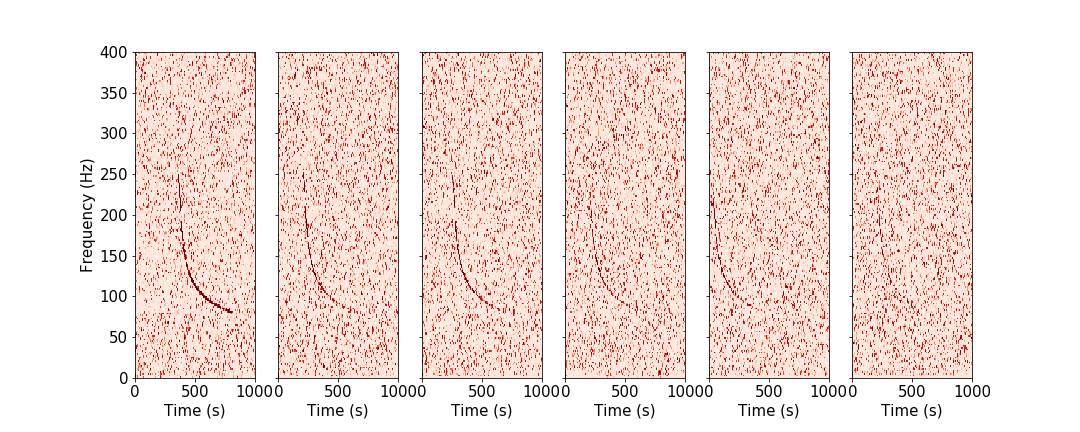}
    \end{tabular}
    \begin{tabular}{c}
        \includegraphics[scale=0.53, trim={2.3cm 0.5cm 1.5cm 1cm},clip]{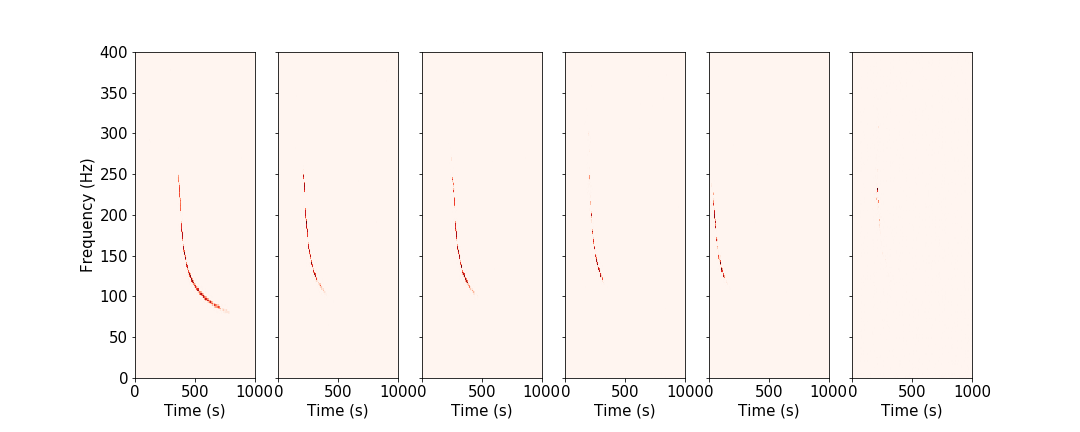}
    \end{tabular}
    \caption{Detection performance of ALBUS on 6 different visibility levels (from left to right : 60, 40, 30, 20, 16 and 12) for the waveform model \textit{GRBplateau}. The top panel shows the input images and the lower panel shows the output of ALBUS.}
    \label{detection visibility 2}
\end{figure*}

The detection performance as a function of the visibility levels can be seen in Figures \ref{detection visibility 1} and \ref{detection visibility 2} for two different long-duration waveforms. The minimal visibility level to which ALBUS can identify most of the signal lies between 16 and 12, which is consistent with the perception of the human eye with no a-priori knowledge of the injected signal.

\subsection{Background analysis}

\begin{figure*}[!htb]
    \centering
    \begin{tabular}{cc}
        \includegraphics[scale=0.4, trim={0.5cm 0.5cm 0.5cm 1cm},clip]{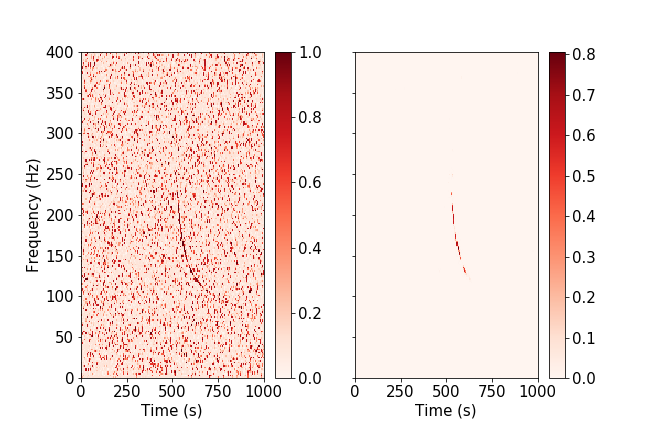} &
        \includegraphics[scale=0.4, trim={0.5cm 0.5cm 0.5cm 1cm},clip]{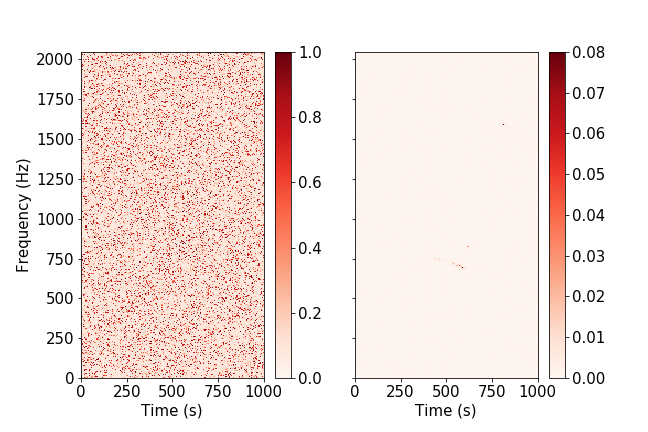}
    \end{tabular}
    \begin{tabular}{cc}
        \includegraphics[scale=0.4, trim={0.5cm 0.5cm 0.5cm 1cm},clip]{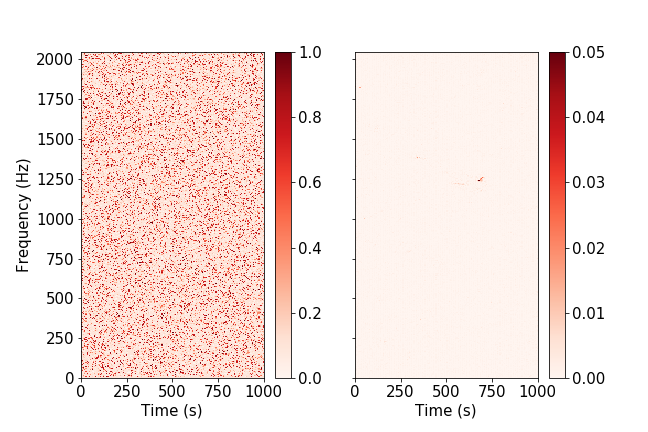} &
        \includegraphics[scale=0.4, trim={0.5cm 0.5cm 0.5cm 1cm},clip]{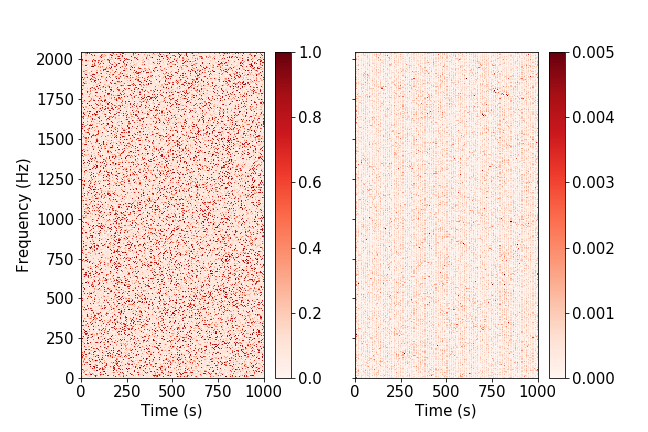}
    \end{tabular}
    \caption{Detection performances of ALBUS on background spectrograms. The top left panel displays the output map for a long-duration signal (\textit{GRBplateau}) for comparison. The left image of each panel is the red channel of the input image and the right panel shows the output of ALBUS.}
    \label{examples background}
\end{figure*}

Figure \ref{examples background} compares the output of ALBUS for 3 different background images to its output when a long-duration waveform is injected. The output map shows correlation values smaller than 0.1. This trend is observed for all the processed spectrograms, with the exception of some isolated hot pixels that can show values up to $0.5$. In any case, the highlighted pixels appear sparse and unconnected for background spectrograms, confirming that our network is searching for connectivity among high value pixels.

\section{Discussion and conclusion}\label{section:conclusion}

We have shown that convolutional neural networks can be applied to the search for minute-long gravitational wave transients in the time-frequency space of the cross-correlated LIGO noise. Our approach allows a fast and pixel-precise identification of the long-duration signals with no training on the latter. The speed of neural networks naturally allows to extend our algorithm to low-latency searches. With training on data accumulated during the first month of the observing run, after which the network would be active for the rest of the run. The whole process including the data acquisition, the spectrogram generation and the forward pass to ALBUS can be carried out in just a few seconds. A low-latency implementation will then consist in repeating this process periodically. \\

In a low-latency implementation, coherent glitches appearing in both Hanford and Livingston interferometers will be detected. There is a need to remove these in order to avoid sending false alarms to other astronomers. Among the low-latency tools to identify and classify glitches, GravitySpy \cite{GravitySpy} and Omicron \cite{Omicron} can produce triggers in few minutes. We could make use of these triggers to discriminate cross-correlated glitches from short signals like ADI and ECBC (see Figure \ref{example waveforms}), that have the potential to be misclassified as a glitch. An alternative is to train a second network to recognize only coherent glitches, appearing as vertical lines, and run it in parallel to the current algorithm. This new network would then serve as a check for glitches. \\

The threshold for the detection of long-duration signals is determined by the highest background candidates, i.e. the background candidates that show the highest detection score as defined by a particular pipeline. Usually, the highest candidates are identified after analysing at least 50 years of background, making more than 1 million spectrograms to process \cite{O3_long_duration_paper}. In order to rank these candidates and automate the detection, a detection statistics needs to be defined in follow-up works.


\section*{Acknowledgements}
The authors thank Jean-René Cudell, Grégory Baltus and Melissa Lopez for useful discussions and comments. V.B. is supported by the Gravitational Wave Science (GWAS) grant funded by the French Community of Belgium. M.F. acknowledges the support of the Fonds de la Recherche Scientifique-FNRS, Belgium, under grant No. 4.4501.19. This material is based upon work supported by NSF's LIGO Laboratory which is a major facility fully funded by the National Science Foundation. The authors are grateful for computational resources provided by the LIGO Laboratory and supported by the National Science Foundation Grants No. PHY-0757058 and No. PHY-0823459. 

\newpage
\bibliographystyle{apsrev}
\bibliography{main.bib}

\begin{thebibliography}{56}
\expandafter\ifx\csname natexlab\endcsname\relax\def\natexlab#1{#1}\fi
\expandafter\ifx\csname bibnamefont\endcsname\relax
  \def\bibnamefont#1{#1}\fi
\expandafter\ifx\csname bibfnamefont\endcsname\relax
  \def\bibfnamefont#1{#1}\fi
\expandafter\ifx\csname citenamefont\endcsname\relax
  \def\citenamefont#1{#1}\fi
\expandafter\ifx\csname url\endcsname\relax
  \def\url#1{\texttt{#1}}\fi
\expandafter\ifx\csname urlprefix\endcsname\relax\def\urlprefix{URL }\fi
\providecommand{\bibinfo}[2]{#2}
\providecommand{\eprint}[2][]{\url{#2}}

\bibitem[{\citenamefont{Aasi et~al.}(2015{\natexlab{a}})}]{aLIGO}
\bibinfo{author}{\bibfnamefont{J.}~\bibnamefont{Aasi}} \bibnamefont{et~al.},
  \bibinfo{journal}{Class. Quant. Grav.} \textbf{\bibinfo{volume}{32}},
  \bibinfo{pages}{074001} (\bibinfo{year}{2015}{\natexlab{a}}), ISSN
  \bibinfo{issn}{1361-6382},
  \urlprefix\url{http://dx.doi.org/10.1088/0264-9381/32/7/074001}.

\bibitem[{\citenamefont{Abbott et~al.}(2016{\natexlab{a}})}]{first_detection}
\bibinfo{author}{\bibfnamefont{B.~P.} \bibnamefont{Abbott}}
  \bibnamefont{et~al.} (\bibinfo{collaboration}{LIGO Scientific Collaboration
  and Virgo Collaboration}), \bibinfo{journal}{Phys. Rev. Lett.}
  \textbf{\bibinfo{volume}{116}}, \bibinfo{pages}{061102}
  (\bibinfo{year}{2016}{\natexlab{a}}),
  \urlprefix\url{https://link.aps.org/doi/10.1103/PhysRevLett.116.061102}.

\bibitem[{\citenamefont{Abbott et~al.}(2017)}]{neutron_star}
\bibinfo{author}{\bibfnamefont{B.~P.} \bibnamefont{Abbott}}
  \bibnamefont{et~al.}, \bibinfo{journal}{Phys. Rev. Lett.}
  \textbf{\bibinfo{volume}{119}} (\bibinfo{year}{2017}), ISSN
  \bibinfo{issn}{1079-7114},
  \urlprefix\url{http://dx.doi.org/10.1103/PhysRevLett.119.161101}.

\bibitem[{\citenamefont{Abbott et~al.}(2021)}]{NSBH}
\bibinfo{author}{\bibfnamefont{R.}~\bibnamefont{Abbott}} \bibnamefont{et~al.},
  \bibinfo{journal}{The Astrophysical Journal Letters}
  \textbf{\bibinfo{volume}{915}}, \bibinfo{pages}{L5} (\bibinfo{year}{2021}),
  \urlprefix\url{https://doi.org/10.3847/2041-8213/ac082e}.

\bibitem[{\citenamefont{{The LIGO Scientific Collaboration}
  et~al.}(2021{\natexlab{a}})\citenamefont{{The LIGO Scientific Collaboration},
  {the Virgo Collaboration}, {the KAGRA Collaboration}, {Abbott}, and
  others.}}]{GWTC3}
\bibinfo{author}{\bibnamefont{{The LIGO Scientific Collaboration}}},
  \bibinfo{author}{\bibnamefont{{the Virgo Collaboration}}},
  \bibinfo{author}{\bibnamefont{{the KAGRA Collaboration}}},
  \bibinfo{author}{\bibfnamefont{R.}~\bibnamefont{{Abbott}}}, \bibnamefont{and}
  \bibinfo{author}{\bibnamefont{others.}}, \bibinfo{journal}{arXiv e-prints}
  \bibinfo{eid}{arXiv:2111.03606} (\bibinfo{year}{2021}{\natexlab{a}}),
  \eprint{2111.03606}.

\bibitem[{\citenamefont{Acernese et~al.}(2014)}]{Virgo}
\bibinfo{author}{\bibfnamefont{F.}~\bibnamefont{Acernese}}
  \bibnamefont{et~al.}, \bibinfo{journal}{Class. Quant. Grav.}
  \textbf{\bibinfo{volume}{32}}, \bibinfo{pages}{024001}
  (\bibinfo{year}{2014}), ISSN \bibinfo{issn}{1361-6382},
  \urlprefix\url{http://dx.doi.org/10.1088/0264-9381/32/2/024001}.

\bibitem[{\citenamefont{Van~Putten}(2001)}]{adi}
\bibinfo{author}{\bibfnamefont{M.~H. P.~M.} \bibnamefont{Van~Putten}},
  \bibinfo{journal}{Phys. Rev. Lett.} \textbf{\bibinfo{volume}{87}},
  \bibinfo{pages}{091101} (\bibinfo{year}{2001}).

\bibitem[{\citenamefont{Dall'Osso et~al.}(2015)\citenamefont{Dall'Osso,
  Giacomazzo, Perna, and Stella}}]{magnetars}
\bibinfo{author}{\bibfnamefont{S.}~\bibnamefont{Dall'Osso}},
  \bibinfo{author}{\bibfnamefont{B.}~\bibnamefont{Giacomazzo}},
  \bibinfo{author}{\bibfnamefont{R.}~\bibnamefont{Perna}}, \bibnamefont{and}
  \bibinfo{author}{\bibfnamefont{L.}~\bibnamefont{Stella}},
  \bibinfo{journal}{\apj} \textbf{\bibinfo{volume}{798}}, \bibinfo{pages}{25}
  (\bibinfo{year}{2015}).

\bibitem[{\citenamefont{Van~Putten et~al.}(2019)\citenamefont{Van~Putten,
  Levinson, Frontera, Guidorzi, Amati, and Della~Valle}}]{supernovae}
\bibinfo{author}{\bibfnamefont{M.~H. P.~M.} \bibnamefont{Van~Putten}},
  \bibinfo{author}{\bibfnamefont{A.}~\bibnamefont{Levinson}},
  \bibinfo{author}{\bibfnamefont{F.}~\bibnamefont{Frontera}},
  \bibinfo{author}{\bibfnamefont{C.}~\bibnamefont{Guidorzi}},
  \bibinfo{author}{\bibfnamefont{L.}~\bibnamefont{Amati}}, \bibnamefont{and}
  \bibinfo{author}{\bibfnamefont{M.}~\bibnamefont{Della~Valle}},
  \bibinfo{journal}{Eur. Phys. J. Plus} \textbf{\bibinfo{volume}{134}},
  \bibinfo{pages}{537} (\bibinfo{year}{2019}), \eprint{1709.04455}.

\bibitem[{\citenamefont{Corsi and M\'esz\'aros}(2009)}]{GRB}
\bibinfo{author}{\bibfnamefont{A.}~\bibnamefont{Corsi}} \bibnamefont{and}
  \bibinfo{author}{\bibfnamefont{P.}~\bibnamefont{M\'esz\'aros}},
  \bibinfo{journal}{The Astrophysical Journal} \textbf{\bibinfo{volume}{702}},
  \bibinfo{pages}{1171} (\bibinfo{year}{2009}).

\bibitem[{\citenamefont{Piro and Thrane}(2012)}]{fallback_accretion_NS}
\bibinfo{author}{\bibfnamefont{A.}~\bibnamefont{Piro}} \bibnamefont{and}
  \bibinfo{author}{\bibfnamefont{E.}~\bibnamefont{Thrane}},
  \bibinfo{journal}{The Astrophysical Journal} \textbf{\bibinfo{volume}{761}},
  \bibinfo{pages}{63} (\bibinfo{year}{2012}).

\bibitem[{\citenamefont{Klimenko et~al.}(2016)\citenamefont{Klimenko, Vedovato,
  Drago, Salemi, Tiwari, Prodi, Lazzaro, Ackley, Tiwari, Da~Silva
  et~al.}}]{cWB}
\bibinfo{author}{\bibfnamefont{S.}~\bibnamefont{Klimenko}},
  \bibinfo{author}{\bibfnamefont{G.}~\bibnamefont{Vedovato}},
  \bibinfo{author}{\bibfnamefont{M.}~\bibnamefont{Drago}},
  \bibinfo{author}{\bibfnamefont{F.}~\bibnamefont{Salemi}},
  \bibinfo{author}{\bibfnamefont{V.}~\bibnamefont{Tiwari}},
  \bibinfo{author}{\bibfnamefont{G.~A.} \bibnamefont{Prodi}},
  \bibinfo{author}{\bibfnamefont{C.}~\bibnamefont{Lazzaro}},
  \bibinfo{author}{\bibfnamefont{K.}~\bibnamefont{Ackley}},
  \bibinfo{author}{\bibfnamefont{S.}~\bibnamefont{Tiwari}},
  \bibinfo{author}{\bibfnamefont{C.~F.} \bibnamefont{Da~Silva}},
  \bibnamefont{et~al.}, \bibinfo{journal}{Phys. Rev. D}
  \textbf{\bibinfo{volume}{93}}, \bibinfo{pages}{042004}
  (\bibinfo{year}{2016}).

\bibitem[{\citenamefont{Thrane and Coughlin}(2015)}]{STAMP_1}
\bibinfo{author}{\bibfnamefont{E.}~\bibnamefont{Thrane}} \bibnamefont{and}
  \bibinfo{author}{\bibfnamefont{M.}~\bibnamefont{Coughlin}},
  \bibinfo{journal}{Phys. Rev. Lett.} \textbf{\bibinfo{volume}{115}},
  \bibinfo{pages}{181102} (\bibinfo{year}{2015}).

\bibitem[{\citenamefont{Prestegard}(2016)}]{STAMP_2}
\bibinfo{author}{\bibfnamefont{T.}~\bibnamefont{Prestegard}},
  \bibinfo{journal}{University of Minnesota Thesis}  (\bibinfo{year}{2016}),
  \bibinfo{note}{\url{https://hdl.handle.net/11299/182183}}.

\bibitem[{\citenamefont{Macquet et~al.}(2021)\citenamefont{Macquet, Bizouard,
  Christensen, and Coughlin}}]{PySTAMPAS}
\bibinfo{author}{\bibfnamefont{A.}~\bibnamefont{Macquet}},
  \bibinfo{author}{\bibfnamefont{M.~A.} \bibnamefont{Bizouard}},
  \bibinfo{author}{\bibfnamefont{N.}~\bibnamefont{Christensen}},
  \bibnamefont{and} \bibinfo{author}{\bibfnamefont{M.}~\bibnamefont{Coughlin}},
  \bibinfo{journal}{Physical Review D} \textbf{\bibinfo{volume}{104}}
  (\bibinfo{year}{2021}), ISSN \bibinfo{issn}{2470-0029},
  \urlprefix\url{http://dx.doi.org/10.1103/PhysRevD.104.102005}.

\bibitem[{\citenamefont{Coyne et~al.}(2016)\citenamefont{Coyne, Corsi, and
  Owen}}]{cocoA}
\bibinfo{author}{\bibfnamefont{R.}~\bibnamefont{Coyne}},
  \bibinfo{author}{\bibfnamefont{A.}~\bibnamefont{Corsi}}, \bibnamefont{and}
  \bibinfo{author}{\bibfnamefont{B.~J.} \bibnamefont{Owen}},
  \bibinfo{journal}{Physical Review D} \textbf{\bibinfo{volume}{93}}
  (\bibinfo{year}{2016}), ISSN \bibinfo{issn}{2470-0029},
  \urlprefix\url{http://dx.doi.org/10.1103/PhysRevD.93.104059}.

\bibitem[{\citenamefont{Fays}(2017)}]{Fays}
\bibinfo{author}{\bibfnamefont{M.}~\bibnamefont{Fays}},
  \bibinfo{journal}{Cardiff University Thesis}  (\bibinfo{year}{2017}),
  \bibinfo{note}{\url{http://orca.cardiff.ac.uk/id/eprint/110245}}.

\bibitem[{\citenamefont{Flanagan and Hughes}(2005)}]{basics_GW}
\bibinfo{author}{\bibfnamefont{E.~E.} \bibnamefont{Flanagan}} \bibnamefont{and}
  \bibinfo{author}{\bibfnamefont{S.~A.} \bibnamefont{Hughes}},
  \bibinfo{journal}{New Journal of Physics} \textbf{\bibinfo{volume}{7}},
  \bibinfo{pages}{204–204} (\bibinfo{year}{2005}), ISSN
  \bibinfo{issn}{1367-2630},
  \urlprefix\url{http://dx.doi.org/10.1088/1367-2630/7/1/204}.

\bibitem[{\citenamefont{Welch}(1967)}]{Welch}
\bibinfo{author}{\bibfnamefont{P.}~\bibnamefont{Welch}}, \bibinfo{journal}{IEEE
  Transactions on Audio and Electroacoustics} \textbf{\bibinfo{volume}{15}},
  \bibinfo{pages}{70} (\bibinfo{year}{1967}).

\bibitem[{\citenamefont{Was et~al.}(2009)\citenamefont{Was, Bizouard, Brisson,
  Cavalier, Davier, Hello, Leroy, Robinet, and Vavoulidis}}]{time_slides}
\bibinfo{author}{\bibfnamefont{M.}~\bibnamefont{Was}},
  \bibinfo{author}{\bibfnamefont{M.~A.} \bibnamefont{Bizouard}},
  \bibinfo{author}{\bibfnamefont{V.}~\bibnamefont{Brisson}},
  \bibinfo{author}{\bibfnamefont{F.}~\bibnamefont{Cavalier}},
  \bibinfo{author}{\bibfnamefont{M.}~\bibnamefont{Davier}},
  \bibinfo{author}{\bibfnamefont{P.}~\bibnamefont{Hello}},
  \bibinfo{author}{\bibfnamefont{N.}~\bibnamefont{Leroy}},
  \bibinfo{author}{\bibfnamefont{F.}~\bibnamefont{Robinet}}, \bibnamefont{and}
  \bibinfo{author}{\bibfnamefont{M.}~\bibnamefont{Vavoulidis}},
  \bibinfo{journal}{Class. Quant. Grav.} \textbf{\bibinfo{volume}{27}},
  \bibinfo{pages}{015005} (\bibinfo{year}{2009}),
  \urlprefix\url{https://doi.org/10.1088/0264-9381/27/1/015005}.

\bibitem[{\citenamefont{Davis et~al.}(2021)}]{detchar2021}
\bibinfo{author}{\bibfnamefont{D.}~\bibnamefont{Davis}} \bibnamefont{et~al.},
  \bibinfo{journal}{Class. Quant. Grav.} \textbf{\bibinfo{volume}{38}},
  \bibinfo{pages}{135014} (\bibinfo{year}{2021}),
  \urlprefix\url{https://doi.org/10.1088/1361-6382/abfd85}.

\bibitem[{\citenamefont{Bengio}(2013)}]{bengio_recommendations}
\bibinfo{author}{\bibfnamefont{Y.}~\bibnamefont{Bengio}}, in
  \emph{\bibinfo{booktitle}{Neural Networks: Tricks of the Trade}}
  (\bibinfo{publisher}{Springer}, \bibinfo{year}{2013}).

\bibitem[{\citenamefont{Barrett et~al.}(2005)\citenamefont{Barrett, Hunter,
  Miller, Hsu, and Greenfield}}]{matplotlib}
\bibinfo{author}{\bibfnamefont{P.}~\bibnamefont{Barrett}},
  \bibinfo{author}{\bibfnamefont{J.}~\bibnamefont{Hunter}},
  \bibinfo{author}{\bibfnamefont{J.}~\bibnamefont{Miller}},
  \bibinfo{author}{\bibfnamefont{J.-C.} \bibnamefont{Hsu}}, \bibnamefont{and}
  \bibinfo{author}{\bibfnamefont{P.}~\bibnamefont{Greenfield}},
  \bibinfo{journal}{Astronomical Data Analysis Software and Systems XIV ASP
  Conference Series} \textbf{\bibinfo{volume}{347}} (\bibinfo{year}{2005}).

\bibitem[{\citenamefont{Harris et~al.}(2020)}]{Numpy}
\bibinfo{author}{\bibfnamefont{C.~R.} \bibnamefont{Harris}}
  \bibnamefont{et~al.}, \bibinfo{journal}{Nature}
  \textbf{\bibinfo{volume}{585}}, \bibinfo{pages}{357} (\bibinfo{year}{2020}),
  \urlprefix\url{https://doi.org/10.1038/s41586-020-2649-2}.

\bibitem[{\citenamefont{Redmon et~al.}(2016)\citenamefont{Redmon, Divvala,
  Girshick, and Farhadi}}]{YOLO}
\bibinfo{author}{\bibfnamefont{J.}~\bibnamefont{Redmon}},
  \bibinfo{author}{\bibfnamefont{S.}~\bibnamefont{Divvala}},
  \bibinfo{author}{\bibfnamefont{R.}~\bibnamefont{Girshick}}, \bibnamefont{and}
  \bibinfo{author}{\bibfnamefont{A.}~\bibnamefont{Farhadi}}, in
  \emph{\bibinfo{booktitle}{2016 IEEE Conference on Computer Vision and Pattern
  Recognition (CVPR)}} (\bibinfo{year}{2016}), pp. \bibinfo{pages}{779--788}.

\bibitem[{\citenamefont{Tan and Le}(2019)}]{EfficientNet}
\bibinfo{author}{\bibfnamefont{M.}~\bibnamefont{Tan}} \bibnamefont{and}
  \bibinfo{author}{\bibfnamefont{Q.~V.} \bibnamefont{Le}}, in
  \emph{\bibinfo{booktitle}{Proceedings of the 36th International Conference on
  Machine Learning}} (\bibinfo{year}{2019}), pp. \bibinfo{pages}{6105--6114},
  \eprint{1905.11946}.

\bibitem[{\citenamefont{Astone et~al.}(2018)\citenamefont{Astone,
  Cerdá-Durán, Di~Palma, Drago, Muciaccia, Palomba, and Ricci}}]{CCSN}
\bibinfo{author}{\bibfnamefont{P.}~\bibnamefont{Astone}},
  \bibinfo{author}{\bibfnamefont{P.}~\bibnamefont{Cerdá-Durán}},
  \bibinfo{author}{\bibfnamefont{I.}~\bibnamefont{Di~Palma}},
  \bibinfo{author}{\bibfnamefont{M.}~\bibnamefont{Drago}},
  \bibinfo{author}{\bibfnamefont{F.}~\bibnamefont{Muciaccia}},
  \bibinfo{author}{\bibfnamefont{C.}~\bibnamefont{Palomba}}, \bibnamefont{and}
  \bibinfo{author}{\bibfnamefont{F.}~\bibnamefont{Ricci}},
  \bibinfo{journal}{Physical Review D} \textbf{\bibinfo{volume}{98}}
  (\bibinfo{year}{2018}), ISSN \bibinfo{issn}{2470-0029},
  \urlprefix\url{http://dx.doi.org/10.1103/PhysRevD.98.122002}.

\bibitem[{\citenamefont{Corsi and Mészáros}(2009)}]{GRBplateauShort}
\bibinfo{author}{\bibfnamefont{A.}~\bibnamefont{Corsi}} \bibnamefont{and}
  \bibinfo{author}{\bibfnamefont{P.}~\bibnamefont{Mészáros}},
  \bibinfo{journal}{The Astrophysical Journal} \textbf{\bibinfo{volume}{702}},
  \bibinfo{pages}{1171–1178} (\bibinfo{year}{2009}), ISSN
  \bibinfo{issn}{1538-4357},
  \urlprefix\url{http://dx.doi.org/10.1088/0004-637X/702/2/1171}.

\bibitem[{\citenamefont{Zhu and Wu}(2004)}]{class_noise}
\bibinfo{author}{\bibfnamefont{X.}~\bibnamefont{Zhu}} \bibnamefont{and}
  \bibinfo{author}{\bibfnamefont{X.}~\bibnamefont{Wu}},
  \bibinfo{journal}{Artificial Intelligence Review}
  \textbf{\bibinfo{volume}{22}}, \bibinfo{pages}{177} (\bibinfo{year}{2004}).

\bibitem[{\citenamefont{Bengio et~al.}(2009)\citenamefont{Bengio, Louradour,
  Collobert, and Weston}}]{curriculum_learning}
\bibinfo{author}{\bibfnamefont{Y.}~\bibnamefont{Bengio}},
  \bibinfo{author}{\bibfnamefont{J.}~\bibnamefont{Louradour}},
  \bibinfo{author}{\bibfnamefont{R.}~\bibnamefont{Collobert}},
  \bibnamefont{and} \bibinfo{author}{\bibfnamefont{J.}~\bibnamefont{Weston}},
  in \emph{\bibinfo{booktitle}{Proceedings of the 26th Annual International
  Conference on Machine Learning}} (\bibinfo{publisher}{Association for
  Computing Machinery}, \bibinfo{address}{New York, NY, USA},
  \bibinfo{year}{2009}), ICML '09, p. \bibinfo{pages}{41–48}, ISBN
  \bibinfo{isbn}{9781605585161},
  \urlprefix\url{https://doi.org/10.1145/1553374.1553380}.

\bibitem[{\citenamefont{Krizhevsky et~al.}(2012)\citenamefont{Krizhevsky,
  Sutskever, and Hinton}}]{AlexNet}
\bibinfo{author}{\bibfnamefont{A.}~\bibnamefont{Krizhevsky}},
  \bibinfo{author}{\bibfnamefont{I.}~\bibnamefont{Sutskever}},
  \bibnamefont{and} \bibinfo{author}{\bibfnamefont{G.~E.}
  \bibnamefont{Hinton}}, in \emph{\bibinfo{booktitle}{Proceedings of the 25th
  International Conference on Neural Information Processing Systems - Volume
  1}} (\bibinfo{publisher}{Curran Associates Inc.}, \bibinfo{address}{Red Hook,
  NY, USA}, \bibinfo{year}{2012}), NIPS'12, p. \bibinfo{pages}{1097–1105}.

\bibitem[{\citenamefont{Szegedy et~al.}(2015)\citenamefont{Szegedy, Wei,
  Yangqing, Sermanet, Reed, Anguelov, Erhan, Vanhoucke, and
  Rabinovich}}]{GoogleNet}
\bibinfo{author}{\bibfnamefont{C.}~\bibnamefont{Szegedy}},
  \bibinfo{author}{\bibfnamefont{L.}~\bibnamefont{Wei}},
  \bibinfo{author}{\bibfnamefont{J.}~\bibnamefont{Yangqing}},
  \bibinfo{author}{\bibfnamefont{P.}~\bibnamefont{Sermanet}},
  \bibinfo{author}{\bibfnamefont{S.}~\bibnamefont{Reed}},
  \bibinfo{author}{\bibfnamefont{D.}~\bibnamefont{Anguelov}},
  \bibinfo{author}{\bibfnamefont{D.}~\bibnamefont{Erhan}},
  \bibinfo{author}{\bibfnamefont{V.}~\bibnamefont{Vanhoucke}},
  \bibnamefont{and}
  \bibinfo{author}{\bibfnamefont{A.}~\bibnamefont{Rabinovich}}, in
  \emph{\bibinfo{booktitle}{2015 IEEE Conference on Computer Vision and Pattern
  Recognition (CVPR)}} (\bibinfo{year}{2015}), pp. \bibinfo{pages}{1--9}.

\bibitem[{\citenamefont{Virtanen et~al.}(2020)\citenamefont{Virtanen, Gommers,
  Oliphant, Haberland, Reddy, Cournapeau, Burovski, Peterson, Weckesser, Bright
  et~al.}}]{scipy}
\bibinfo{author}{\bibfnamefont{P.}~\bibnamefont{Virtanen}},
  \bibinfo{author}{\bibfnamefont{R.}~\bibnamefont{Gommers}},
  \bibinfo{author}{\bibfnamefont{T.~E.} \bibnamefont{Oliphant}},
  \bibinfo{author}{\bibfnamefont{M.}~\bibnamefont{Haberland}},
  \bibinfo{author}{\bibfnamefont{T.}~\bibnamefont{Reddy}},
  \bibinfo{author}{\bibfnamefont{D.}~\bibnamefont{Cournapeau}},
  \bibinfo{author}{\bibfnamefont{E.}~\bibnamefont{Burovski}},
  \bibinfo{author}{\bibfnamefont{P.}~\bibnamefont{Peterson}},
  \bibinfo{author}{\bibfnamefont{W.}~\bibnamefont{Weckesser}},
  \bibinfo{author}{\bibfnamefont{J.}~\bibnamefont{Bright}},
  \bibnamefont{et~al.}, \bibinfo{journal}{Nature Methods}
  \textbf{\bibinfo{volume}{17}}, \bibinfo{pages}{261–272}
  (\bibinfo{year}{2020}), ISSN \bibinfo{issn}{1548-7105},
  \urlprefix\url{http://dx.doi.org/10.1038/s41592-019-0686-2}.

\bibitem[{\citenamefont{Kaiser}(1966)}]{kaiser_window}
\bibinfo{author}{\bibfnamefont{J.~F.} \bibnamefont{Kaiser}},
  \bibinfo{journal}{Systems Analysis by Digital Computer} pp.
  \bibinfo{pages}{218--285} (\bibinfo{year}{1966}).

\bibitem[{\citenamefont{{The LIGO Scientific Collaboration}
  et~al.}(2021{\natexlab{b}})\citenamefont{{The LIGO Scientific Collaboration},
  {the Virgo Collaboration}, {the KAGRA Collaboration}, Abbott
  et~al.}}]{O3_long_duration_paper}
\bibinfo{author}{\bibnamefont{{The LIGO Scientific Collaboration}}},
  \bibinfo{author}{\bibnamefont{{the Virgo Collaboration}}},
  \bibinfo{author}{\bibnamefont{{the KAGRA Collaboration}}},
  \bibinfo{author}{\bibfnamefont{R.}~\bibnamefont{Abbott}},
  \bibnamefont{et~al.}, \bibinfo{journal}{Phys. Rev. D}
  (\bibinfo{year}{2021}{\natexlab{b}}), \eprint{2107.13796}.

\bibitem[{\citenamefont{Van~Putten}(2016)}]{ISCOchirp}
\bibinfo{author}{\bibfnamefont{M.~H. P.~M.} \bibnamefont{Van~Putten}},
  \bibinfo{journal}{The Astrophysical Journal} \textbf{\bibinfo{volume}{819}},
  \bibinfo{pages}{169} (\bibinfo{year}{2016}), ISSN \bibinfo{issn}{1538-4357},
  \urlprefix\url{http://dx.doi.org/10.3847/0004-637X/819/2/169}.

\bibitem[{\citenamefont{Huerta et~al.}(2017)\citenamefont{Huerta, Kumar,
  Agarwal, George, Schive, Pfeiffer, Haas, Ren, Chu, Boyle et~al.}}]{ecbc}
\bibinfo{author}{\bibfnamefont{E.~A.} \bibnamefont{Huerta}},
  \bibinfo{author}{\bibfnamefont{P.}~\bibnamefont{Kumar}},
  \bibinfo{author}{\bibfnamefont{B.}~\bibnamefont{Agarwal}},
  \bibinfo{author}{\bibfnamefont{D.}~\bibnamefont{George}},
  \bibinfo{author}{\bibfnamefont{H.-Y.} \bibnamefont{Schive}},
  \bibinfo{author}{\bibfnamefont{H.~P.} \bibnamefont{Pfeiffer}},
  \bibinfo{author}{\bibfnamefont{R.}~\bibnamefont{Haas}},
  \bibinfo{author}{\bibfnamefont{W.}~\bibnamefont{Ren}},
  \bibinfo{author}{\bibfnamefont{T.}~\bibnamefont{Chu}},
  \bibinfo{author}{\bibfnamefont{M.}~\bibnamefont{Boyle}},
  \bibnamefont{et~al.}, \bibinfo{journal}{Phys. Rev. D}
  \textbf{\bibinfo{volume}{95}}, \bibinfo{pages}{024038}
  (\bibinfo{year}{2017}),
  \urlprefix\url{https://link.aps.org/doi/10.1103/PhysRevD.95.024038}.

\bibitem[{\citenamefont{Van~Putten}(2002)}]{vanputten}
\bibinfo{author}{\bibfnamefont{M.~H. P.~M.} \bibnamefont{Van~Putten}},
  \bibinfo{journal}{The Astrophysical Journal} \textbf{\bibinfo{volume}{575}},
  \bibinfo{pages}{L71–L74} (\bibinfo{year}{2002}), ISSN
  \bibinfo{issn}{1538-4357}, \urlprefix\url{http://dx.doi.org/10.1086/342781}.

\bibitem[{\citenamefont{Cuoco et~al.}(2020)}]{MLreview}
\bibinfo{author}{\bibfnamefont{E.}~\bibnamefont{Cuoco}} \bibnamefont{et~al.},
  \bibinfo{journal}{Machine Learning: Science and Technology}
  \textbf{\bibinfo{volume}{2}}, \bibinfo{pages}{011002} (\bibinfo{year}{2020}),
  \urlprefix\url{https://doi.org/10.1088/2632-2153/abb93a}.

\bibitem[{\citenamefont{Goodfellow et~al.}(2016)\citenamefont{Goodfellow,
  Bengio, and Courville}}]{CNN}
\bibinfo{author}{\bibfnamefont{I.}~\bibnamefont{Goodfellow}},
  \bibinfo{author}{\bibfnamefont{Y.}~\bibnamefont{Bengio}}, \bibnamefont{and}
  \bibinfo{author}{\bibfnamefont{A.}~\bibnamefont{Courville}},
  \emph{\bibinfo{title}{Deep Learning}} (\bibinfo{publisher}{MIT Press},
  \bibinfo{year}{2016}), \bibinfo{note}{\url{http://www.deeplearningbook.org}}.

\bibitem[{\citenamefont{Menéndez-Vázquez and others.}(2021)}]{CBC_cnn}
\bibinfo{author}{\bibfnamefont{A.}~\bibnamefont{Menéndez-Vázquez}}
  \bibnamefont{and} \bibinfo{author}{\bibnamefont{others.}},
  \bibinfo{journal}{Phys. Rev. D} \textbf{\bibinfo{volume}{103}}
  (\bibinfo{year}{2021}), ISSN \bibinfo{issn}{2470-0029},
  \urlprefix\url{http://dx.doi.org/10.1103/PhysRevD.103.062004}.

\bibitem[{\citenamefont{L\'opez et~al.}(2021)}]{Melissa}
\bibinfo{author}{\bibfnamefont{M.}~\bibnamefont{L\'opez}} \bibnamefont{et~al.},
  \bibinfo{journal}{Phys. Rev. D} \textbf{\bibinfo{volume}{103}},
  \bibinfo{pages}{063011} (\bibinfo{year}{2021}),
  \urlprefix\url{https://link.aps.org/doi/10.1103/PhysRevD.103.063011}.

\bibitem[{\citenamefont{Baltus et~al.}(2021)}]{Baltus}
\bibinfo{author}{\bibfnamefont{G.}~\bibnamefont{Baltus}} \bibnamefont{et~al.},
  \bibinfo{journal}{Phys. Rev. D} \textbf{\bibinfo{volume}{103}},
  \bibinfo{pages}{102003} (\bibinfo{year}{2021}).

\bibitem[{\citenamefont{Krastev}(2020)}]{Krastev2020}
\bibinfo{author}{\bibfnamefont{P.}~\bibnamefont{Krastev}},
  \bibinfo{journal}{Physics Letters B} \textbf{\bibinfo{volume}{803}},
  \bibinfo{pages}{135330} (\bibinfo{year}{2020}).

\bibitem[{\citenamefont{Krastev et~al.}(2021)}]{parameter_estimation}
\bibinfo{author}{\bibfnamefont{P.~G.~K.} \bibnamefont{Krastev}}
  \bibnamefont{et~al.}, \bibinfo{journal}{Physics Letters B}
  \textbf{\bibinfo{volume}{815}}, \bibinfo{pages}{136161}
  (\bibinfo{year}{2021}), ISSN \bibinfo{issn}{0370-2693},
  \urlprefix\url{http://dx.doi.org/10.1016/j.physletb.2021.136161}.

\bibitem[{\citenamefont{Zevin et~al.}(2017)}]{GravitySpy}
\bibinfo{author}{\bibfnamefont{M.}~\bibnamefont{Zevin}} \bibnamefont{et~al.},
  \bibinfo{journal}{Class. Quant. Grav.} \textbf{\bibinfo{volume}{34}},
  \bibinfo{pages}{064003} (\bibinfo{year}{2017}), ISSN
  \bibinfo{issn}{1361-6382},
  \urlprefix\url{http://dx.doi.org/10.1088/1361-6382/aa5cea}.

\bibitem[{\citenamefont{Skliris et~al.}(2020)\citenamefont{Skliris, Norman, and
  Sutton}}]{short_duration_burst}
\bibinfo{author}{\bibfnamefont{V.}~\bibnamefont{Skliris}},
  \bibinfo{author}{\bibfnamefont{M.}~\bibnamefont{Norman}}, \bibnamefont{and}
  \bibinfo{author}{\bibfnamefont{P.}~\bibnamefont{Sutton}}
  (\bibinfo{year}{2020}), \bibinfo{note}{arXiv:2009.14611},
  \eprint{2009.14611}.

\bibitem[{\citenamefont{Goodfellow et~al.}(2014)}]{Original_GAN}
\bibinfo{author}{\bibfnamefont{I.~J.} \bibnamefont{Goodfellow}}
  \bibnamefont{et~al.}, in \emph{\bibinfo{booktitle}{Proceedings of the 27th
  International Conference on Neural Information Processing Systems - Volume
  2}} (\bibinfo{publisher}{MIT Press}, \bibinfo{address}{Cambridge, MA, USA},
  \bibinfo{year}{2014}), NIPS'14, p. \bibinfo{pages}{2672–2680}.

\bibitem[{\citenamefont{McGinn et~al.}(2021)}]{CGAN_bursts}
\bibinfo{author}{\bibfnamefont{J.}~\bibnamefont{McGinn}} \bibnamefont{et~al.},
  \bibinfo{journal}{Class. Quant. Grav.} \textbf{\bibinfo{volume}{38}},
  \bibinfo{pages}{155005} (\bibinfo{year}{2021}), \eprint{2103.01641}.

\bibitem[{\citenamefont{Xing et~al.}(2019)\citenamefont{Xing, Xie, Shi, Chen,
  Zhang, and Yang}}]{albus}
\bibinfo{author}{\bibfnamefont{F.}~\bibnamefont{Xing}},
  \bibinfo{author}{\bibfnamefont{Y.}~\bibnamefont{Xie}},
  \bibinfo{author}{\bibfnamefont{X.}~\bibnamefont{Shi}},
  \bibinfo{author}{\bibfnamefont{P.}~\bibnamefont{Chen}},
  \bibinfo{author}{\bibfnamefont{Z.}~\bibnamefont{Zhang}}, \bibnamefont{and}
  \bibinfo{author}{\bibfnamefont{L.}~\bibnamefont{Yang}}, \bibinfo{journal}{BMC
  Bioinformatics} \textbf{\bibinfo{volume}{20}} (\bibinfo{year}{2019}).

\bibitem[{\citenamefont{Paszke et~al.}(2019)}]{Pytorch}
\bibinfo{author}{\bibfnamefont{A.}~\bibnamefont{Paszke}} \bibnamefont{et~al.},
  in \emph{\bibinfo{booktitle}{Advances in Neural Information Processing
  Systems 32}}, edited by
  \bibinfo{editor}{\bibfnamefont{H.}~\bibnamefont{Wallach}},
  \bibinfo{editor}{\bibfnamefont{H.}~\bibnamefont{Larochelle}},
  \bibinfo{editor}{\bibfnamefont{A.}~\bibnamefont{Beygelzimer}},
  \bibinfo{editor}{\bibfnamefont{F.}~\bibnamefont{d\textquotesingle
  Alch\'{e}-Buc}}, \bibinfo{editor}{\bibfnamefont{E.}~\bibnamefont{Fox}},
  \bibnamefont{and} \bibinfo{editor}{\bibfnamefont{R.}~\bibnamefont{Garnett}}
  (\bibinfo{publisher}{Curran Associates, Inc.}, \bibinfo{year}{2019}), pp.
  \bibinfo{pages}{8024--8035},
  \urlprefix\url{http://papers.neurips.cc/paper/9015-pytorch-an-imperative-style-high-performance-deep-learning-library.pdf}.

\bibitem[{\citenamefont{Kingma and Ba}(2014)}]{Adam}
\bibinfo{author}{\bibfnamefont{D.}~\bibnamefont{Kingma}} \bibnamefont{and}
  \bibinfo{author}{\bibfnamefont{J.}~\bibnamefont{Ba}},
  \bibinfo{journal}{International Conference on Learning Representations}
  (\bibinfo{year}{2014}).

\bibitem[{\citenamefont{{Dall'Osso} et~al.}(2015)\citenamefont{{Dall'Osso},
  {Giacomazzo}, {Perna}, and {Stella}}}]{magnetar_waveform}
\bibinfo{author}{\bibfnamefont{S.}~\bibnamefont{{Dall'Osso}}},
  \bibinfo{author}{\bibfnamefont{B.}~\bibnamefont{{Giacomazzo}}},
  \bibinfo{author}{\bibfnamefont{R.}~\bibnamefont{{Perna}}}, \bibnamefont{and}
  \bibinfo{author}{\bibfnamefont{L.}~\bibnamefont{{Stella}}},
  \bibinfo{journal}{\apj} \textbf{\bibinfo{volume}{798}}, \bibinfo{eid}{25}
  (\bibinfo{year}{2015}), \eprint{1408.0013}.

\bibitem[{\citenamefont{Abbott
  et~al.}(2016{\natexlab{b}})}]{noise_characterization}
\bibinfo{author}{\bibfnamefont{B.~P.} \bibnamefont{Abbott}}
  \bibnamefont{et~al.}, \bibinfo{journal}{Class. and Quant. Grav.}
  \textbf{\bibinfo{volume}{33}}, \bibinfo{pages}{134001}
  (\bibinfo{year}{2016}{\natexlab{b}}).

\bibitem[{\citenamefont{Aasi
  et~al.}(2015{\natexlab{b}})}]{noise_characterization2}
\bibinfo{author}{\bibfnamefont{J.}~\bibnamefont{Aasi}} \bibnamefont{et~al.},
  \bibinfo{journal}{Class. Quant. Grav.} \textbf{\bibinfo{volume}{32}},
  \bibinfo{pages}{115012} (\bibinfo{year}{2015}{\natexlab{b}}).

\bibitem[{\citenamefont{Robinet et~al.}(2020)\citenamefont{Robinet, Arnaud,
  Leroy, Lundgren, Macleod, and McIver}}]{Omicron}
\bibinfo{author}{\bibfnamefont{F.}~\bibnamefont{Robinet}},
  \bibinfo{author}{\bibfnamefont{N.}~\bibnamefont{Arnaud}},
  \bibinfo{author}{\bibfnamefont{N.}~\bibnamefont{Leroy}},
  \bibinfo{author}{\bibfnamefont{A.}~\bibnamefont{Lundgren}},
  \bibinfo{author}{\bibfnamefont{D.}~\bibnamefont{Macleod}}, \bibnamefont{and}
  \bibinfo{author}{\bibfnamefont{J.}~\bibnamefont{McIver}},
  \bibinfo{journal}{SoftwareX} \textbf{\bibinfo{volume}{12}},
  \bibinfo{pages}{100620} (\bibinfo{year}{2020}), ISSN
  \bibinfo{issn}{2352-7110},
  \urlprefix\url{https://www.sciencedirect.com/science/article/pii/S2352711020303332}.

\end{thebibliography}

\end{document}